\documentclass[10pt,twoside]{article}
\usepackage{fancyhdr}
\usepackage{titlesec}
\usepackage{cite}
\usepackage{ifthen}

\titleformat{\section}{\centering\large\bfseries}{\S\arabic{section}}{1em}{}
\newboolean{first}
\setboolean{first}{true}

\begin{document}

\setlength\abovedisplayskip{2pt}
\setlength\abovedisplayshortskip{0pt}
\setlength\belowdisplayskip{2pt}
\setlength\belowdisplayshortskip{0pt}

\title{\bf \Large  New Construction of Authentication Codes with Arbitration from Pseudo-Symplectic Geometry over Finite Fields
\author{Chen Shang-di   \  \ Zhao
Da-wei \\ \small \it College of Science, Civil Aviation University of China,Tianjin,300300, P.R.China }\date{}} \maketitle

 \footnote{MR Subject
Classification: 94A62; 94A60 } \footnote{Supported by the
NNSF of China(60776810), the NSF of Tianjin(08JCYBJC13900).}
\footnote{Address: College of Science, Civil Aviation University of China, Tianjin 300300, PR CHina.}
\footnote{E-mail address: sdchen@cauc.edu.cn}
\begin{center}
\begin{minipage}{135mm}
{\bf \small Abstract}.\hskip 2mm {\small A new construction of authentication codes with arbitration from
pseudo-symplectic geometry over finite fields is given. The
parameters and the probabilities of deceptions of the codes are also
computed.}
\end{minipage}
\end{center}

\thispagestyle{fancyplain} \fancyhead{}
\fancyhead[L]{\textit{My paper}\\
} \fancyfoot{} \vskip 10mm

\setcounter{section}{1} \setcounter{equation}{0}
\renewcommand{\theequation}
{\arabic{equation}}
\begin{center}
{\large\bf\S\bf1\hspace*{3mm}  Introduction}
\end{center}

To solve the distrust problem of the transmitter and the receiver in
the communications system, Simmons introduced a model of
authentication codes with arbitration (see [1]), we write symply ($
{A}^{2} $-code) defined as follows:\

Let $S$, $E_T$,$E_R$ and $M$ be four non-empty finite sets, and
$f:S\times E_T\rightarrow M$ and $g:M\times E_R\rightarrow S\cup
\{reject\}$ be two maps. The six-tuple $(S,E_T,E_R,M;f,g)$ is called
an authentication code with arbitration (${A}^{2} $-code), if\

(1) The maps $f$ and $g$ are surjective;\

(2) For any $m\in M$ and $e_T\in E_T$, if there is an $s\in S$,
satisfying $f(s,e_T)=m$, then such an $s$ is uniquely determined by
the given $m$ and $e_T$;\

(3) $p(e_T,e_R)\neq 0$ and $f(s,e_T)=m$ implies $g(m,e_R)=s$,
otherwise, $g(m,e_R)=\{reject\}$.\

$S$, $E_T$,$E_R$ and $M$ are called the set of source states, the
set of transmitter's encoding rules, the set of receiver's decoding
rules and the set of messages, respectively; $f$ and $g$ are called
the encoding map and decoding map respectively. The cardinals $|S|$,
$|E_T|$,$|E_R|$ and $|M|$ are called the size parameters of the
code.\

In a authentication system that permits arbitration, this model
includes four attendances: the transmitter, the receiver, the
opponent and the arbiter, and includes five attacks:

1) The opponent's impersonation attack: the largest probability of
an opponent's successful impersonation attack is $P_{I}$ . Then

$$P_{I}=\max\limits_{m\in M}\left\{\frac{\mid{e_{R}\in E_{R}|e_{R}\subset m}\mid}{\mid E_{R} \mid}\right\}.$$

2) The opponent's substitution attack: the largest probability of an
opponent's successful substitution attack is $P_{S}$.
Then$$P_{S}=\max\limits_{m\in M}\left\{\frac{\max\limits_{m\neq
m^{'}\in M} \mid {e_{R}\in E_{R}|e_{R}\subset m \ {\rm and} \
e_{R}\subset m^{'}} \mid} {\mid {e_{R}\in E_{R}|e_{R}\subset m}
\mid}\right\}.
$$

3) The transmitter's impersonation attack: the largest probability
of a transmitter's successful impersonation attack is $P_{T}$ . Then
$$P_{T}=\max\limits_{e_{T}\in E_{T}}\left\{\frac{\max\limits_{m \in M, e_{T}\not \in m }\mid\{e_{R}\in E_{R}|e_{R}\subset m
{\rm \ and} \ p\left(e_{R},e_{T}\right)\neq 0\}\mid}{\mid\{e_{R} \in
E_{R}|p(e_{R},e_{T})\neq 0\}\mid}\right\}.$$

4) The receiver's impersonation attack: the largest probability of a
receiver's successful impersonation attack is $P_{R_{0}}$ . Then

$$P_{R_{0}}=\max\limits_{e_{R}\in E_{R}}\left\{\frac{\max\limits_{m\in M
 }\mid\{e_{T}\in E_{T}|e_{T}\subset m {\rm \ and} \
p\left(e_{R},e_{T}\right)\neq 0\}\mid}{\mid\{e_{T} \in
E_{T}|p(e_{R},e_{T})\neq 0\}\mid}\right\}.$$

5) The receiver's substitution attack: the largest probability of a
receiver's successful substitution attack is $P_{R_{1}}$ .Then

$$P_{R_{1}}=\max\limits_{e_{R}\in E_{R},m\in M}\left\{\frac{\max\limits_{m^{'}
\in M
 }\mid\{e_{T}\in E_{T}|e_{T}\subset m ,m^{'}\ {\rm  and} \
p\left(e_{R},e_{T}\right)\neq 0\}\mid}{\mid\{e_{T} \in
E_{T}|e_{T}\subset m \,and \,p(e_{R},e_{T})\neq 0\}\mid}\right\}.$$
{\bf Notes:} $p\left(e_{R},e_{T}\right)\neq 0$ implies that any
information $s$ encoded by $e_{T}$ can be authenticated by $e_{R}.$

In this paper, the $^tP$ denotes the transpose of a matrix $P$.
Some concepts and notations  refer to [2].

\setcounter{section}{2} \setcounter{equation}{0}
\renewcommand{\theequation}
{\arabic{equation}}
\begin{center}
{\large\bf\S\bf2\hspace*{3mm}  Pseudo-Symplectic Geometry}
\end{center}

Let $ F_q$ be the finite field with $q$ elements, where $q$ is a
power of $2$, $n=2\nu + \delta$ and $\delta$=1,2. Let
$$
  K = \left (
  \begin {array}{ccc}
  0 & I^{(\nu)}\\
  I^{(\nu)}& 0
  \end{array} \right ), \,\,\,\,\,  S_{1} = \left (
  \begin {array}{cc}
  K &\\
 & {1}
  \end{array} \right ),\,\,\,\,\,S_{2} = \left (
  \begin {array}{cccc}
   K &&\\
& 0&1\\
&1&1\end{array} \right )
  $$
and $S_\delta$ is an $(2\nu+\delta)\times(2\nu+\delta)$
non-alternate symmetric matrix.\

The pseudo-symplectic group of degree $(2\nu+\delta)$ over $
F_q$ is defined to be the set of matrices $Ps_{2\nu +
\delta}(F_q)=\{T|T S_\delta\ ^{t}T=S_\delta\}$
denoted by $Ps_{2\nu + \delta}(F_q)$.\

Let $F_q^{(2\nu + \delta)}$ be the $(2\nu + \delta)$
-dimensional row vector space over $F_q$. $Ps_{2\nu +
\delta}(F_q)$ has an action on $F_q^{(2\nu +
\delta)}$ defined as follows:
$$
F_q^{(2\nu + \delta)} \times Ps_{2\nu +
\delta}(F_q) \rightarrow F_q^{(2\nu +
\delta)}\ $$ $$ ((x_1,x_2,\dots,x_{2\nu +
\delta}),T)\rightarrow(x_1,x_2,\dots,x_{2\nu + \delta})T
$$
The vector space $F_q^{(2\nu + \delta)}$ together with
this group action is called the pseudo-symplectic space over the
finite field $ F_q$ of characteristic 2.\

Let $P$ be an $m$-dimensional subspace of $F_q^{(2\nu +
\delta)}$, then $PS_\delta\ ^{t}P$ is cogredient to one of the
following three normal forms
$$ M(\,m,\,2s,\,s\,)=\left (
  \begin {array}{ccccc}
  0&I^{(s)}&\\
  I^{(s)}&0&\\
  &&0^{(m-2s)}
  \end{array} \right)$$\\
$$ M(\,m,\,2s+1,\,s\,)=\left (
  \begin {array}{ccccc}
  0&I^{(s)}&\\
  I^{(s)}&0&\\
   &&1\\
  &&&0^{(m-2s-1)}
  \end{array} \right)$$\\
$$ M(\,m,\,2s+2,\,s\,)=\left (
  \begin {array}{ccccc}
  0&I^{(s)}&\\
  I^{(s)}&0&\\
   &&0&1&\\
   &&1&1&\\
  &&&&0^{(m-2s-2)}
  \end{array} \right)$$\\
for some $s$ such that $0\leq s\leq [m/2]$. We say that $P$ ia a
subspace of type $(m,2s+\tau,s,\epsilon)$, where $\tau$ =0,1 or 2
and $\epsilon$ =0 or 1, if \

(i) $P S_\delta\ ^{t}P$ is cogredient to $M(m,2s+\tau,s)$, and\

 (ii) $e_{2\nu +1}\notin P$ or $e_{2\nu +1}\in P$ according to $\epsilon=0$ or
 $\epsilon=1$, respectively.\

Let $P$ be an $m$-dimensional subspace of $F_q^{(2\nu +
\delta)}$. Denote by $P^\perp $ the set of vectors which are
orthogonal to every vector of $P$, i.e.,
$$ P^\perp=\{y\in F_q^{(2\nu +
\delta)}|yS_\delta\ ^{t}x=0\, for\,\, all\, x\in P\}
$$Obviously, $P^\perp $ is a $(2\nu +\delta-m)$-dimensional subspace of
$F_q^{(2\nu + \delta)}$ .

More properties of geometry of pseudo-symplectic  groups over
finite fields of characteristic $2$ can be found in [2].

In [3-5] several constructions of authentication codes with
arbitration from the geometry of classical groups over finite fields
were given and studied. In this paper a construction of
authentication codes with arbitration from pseudo-symplectic
geometry over finite fields is given. The parameters and the
probabilities of deceptions of the codes are also computed.

\setcounter{section}{3} \setcounter{equation}{0}
\renewcommand{\theequation}
{\arabic{equation}}
\begin{center}
{\large\bf\S\bf3\hspace*{3mm}  Construction}
\end{center}
\par\noindent
Assume that $n=(2\nu +\delta),\, s-1\leq s_0\leq\nu, \,2s\leq m_0,
2s_0\leq m_0$. Let $ \langle\nu_0, e_{2\nu +1} \rangle$ be a fixed
subspace of type (2,0,0,1) in the $(2\nu +2)$-dimensional
pseudo-symplectic space $F_q^{(2\nu +2)}$; $P_0$ is a
fixed subspace of type $(m_0,2s_0,s_0,1)$ in  $F_q^{(2\nu
+2)}$ and $ \langle\nu_0, e_{2\nu +1} \rangle\subset
P_0\subset{\langle\nu_0, e_{2\nu +1} \rangle}^\bot $. The set of
source states $S=\{s|s$ is a subspace of type \,$(2s,
2(s-1),s-1,1)$\, and\, $ \langle\nu_0, e_{2\nu +1} \rangle\subset
s\subset P_0$\}; \,the set of transmitter's encoding rules
$E_T$=$\{e_T |e_T$\, is a subspace of type\, (4,4,1,1)\, and
\,$e_T\cap P_0 = \langle\nu_0, e_{2\nu +1} \rangle$\}; the set of
receiver's decoding rules $E_R$=$\{e_R |e_R$\, is a subspace of
type\, (2,2,0,1) in the $(2\nu +2)$-dimensional pseudo-symplectic
space $F_q^{(2\nu +2)}\}$; the set of messages $M=\{m|m$
is a subspace of type \,$(2s+2, 2s+2,s,1)$, $ \langle\nu_0, e_{2\nu
+1} \rangle\subset m$, and $m\cap P_0$ is a subspace of type $(2s,
2(s-1),s-1,1)$\}.\

Define the encoding map:
$$
f:S\times E_T\rightarrow M, \,\,(s,e_T)\longmapsto m=s+e_T
$$
and the decoding map:
$$
g:M\times E_R\rightarrow S\cup\{reject\}$$

$$(m,e_{R})\mapsto\left\{\begin{array}{ll}s &$
 if $ e_{R}\subset m, $ where $ s=m\cap P_0.\\ \{reject\}&$ if $ e_{R} \not\subset m.\end{array}\right.
 $$

{\bf Lemma 1.} \ The six-tuple$(S,E_T,E_R,M;f,g)$ is an
authentication code with arbitration, that is

(1) $s+e_T=m\in M$, for all $s\in S$ and $e_T\in E_T$;

(2) for any $m\in M$, $s=m\cap P_0$ is the uniquely source state
contained in $m$ and there is $e_T\in E_T$, such that $m=s+e_T$. \

{\bf Proof. } (1) For any $s\in S$, $s$ is a subspace of type $(2s,
2(s-1),s-1,1)$ and $ \langle\nu_0, e_{2\nu +1} \rangle\subset
S\subset P_0$,\, we can assume that
$$ s=\left(\begin{array}{c}Q\\ \nu_0\\ e_{2\nu+1}\end{array}\right)\begin{array}{c}2s-2\\ 1\\ 1\end{array},$$
then
$$\left(\begin{array}{c}Q\\ \nu_0\\ e_{2\nu+1}\end{array}\right)S_2 \ \begin{array}{l}^t\\ \\ \\  \end{array}\left(\begin{array}{c}Q\\ \nu_0\\ e_{2\nu+1}\end{array}\right)
=\begin{array}{ccccc}\left(\begin{array}{c}0\\ I^{(s-1)}\\ 0\\ 0\end{array}\right.& \begin{array}{c}0\\ 0\\ I^{(s-1)} \\ 0\end{array}&
\begin{array}{c}0\\ 0\\ 0\\ 0\end{array}&\left.\begin{array}{c}0\\ 0\\ 0\\ 0\end{array}\right)\\ s-1&s-1&1&1&\end{array}.$$

For any $e_T\in E_T$, $e_T$ is a subspace of type\,
(4,4,1,1)\, and \,$e_T\cap P_0 = \langle\nu_0, e_{2\nu +1}
\rangle$,\, we can assume that
$$\begin {array}[t] {cc@{\extracolsep{0.2ex}}c}e_T=&\left (
  \begin {array}{ccccc}
 v_0\\
  e_{2\nu +1}\\
  u_1\\
  u_2
  \end{array} \right)&\begin {array}{l}
  \scriptstyle 1\\\scriptstyle 1\\\scriptstyle 1\\\scriptstyle 1
  \end{array}
  \end{array},$$
  then

$$
\left (
  \begin {array}{ccccc}
   v_0\\
  e_{2\nu +1}\\
  u_1\\
  u_2
  \end{array} \right)
  S_2
\ \begin{array}{l}^t\\ \\ \\ \\ \end{array}\left(
\begin {array}{ccccc}
  v_0\\
  e_{2\nu +1}\\
  u_1\\
  u_2
  \end{array}\right)=
  \left (
  \begin {array}{ccccc}
  0&1&0&0\\
  1&0&0&0\\
  0&0&0&1\\
  0&0&1&1
  \end{array} \right) $$

  Obviously, $u_1,u_2\notin S$. Hence $m=s+e_T$ is a
$(2s+2)$-dimensional subspace and $m\cap P_0=s$ is a subspace of
type $(2s, 2(s-1),s-1,1)$. We also have
$$
\begin {array}[t] {cc@{\extracolsep{0.2ex}}c}mS_2\ ^t m=&\left (
  \begin {array}{ccccc}
Q\\
   v_0\\
  e_{2\nu +1}\\
  u_1\\
  u_2
  \end{array} \right)
  \end{array}S_2
{\begin {array}[t] {cc@{\extracolsep{0.2ex}}c}\
\begin{array}{l}^t\\ \\ \\ \\ \end{array}\left (
  \begin {array}{ccccc}
  Q\\
  v_0\\
  e_{2\nu +1}\\
  u_1\\
  u_2
  \end{array} \right)
  \end{array}}=
  \begin {array}[t] {cc@{\extracolsep{0.2ex}}c}&\left (
  \begin {array}{cccccccc}
  0& I^{(s-1)}&0&0& \ast &  \ast   \\
  I^{(s-1)}&0&0&0& \ast &  \ast\\
  0&0&0&1&0&0\\
  0&0&1&0&0&0\\
  \ast &  \ast&0&0&0&1\\
  \ast &  \ast&0&0&1&1
  \end{array} \right)
  \end{array}.
$$
Therefore, $m$ is a subspace of type $(2s+2, 2s+2,s,1)$, $
\langle\nu_0, e_{2\nu +1} \rangle\subset m$, and $m\cap P_0$ is a
subspace of type $(2s, 2(s-1),s-1,1)$, i.e., $m\in M$ is a message.\

(2) If $m\in M$, let $s=m\cap P_0$, then $s$ is a subspace of type
 $(2s, 2(s-1),s-1,1)$  and $ \langle\nu_0, e_{2\nu +1}
\rangle\subset S\subset P_0$, i.e., $s\in S$ is a source state. Now
let
\begin{center} $
\begin {array}[t] {cc@{\extracolsep{0.2ex}}c}s=&\left (
  \begin {array}{ccccc}
  Q\\
  v_0\\
  e_{2\nu +1}
  \end{array} \right)&\begin {array}{l}
  \scriptstyle 2s-2\\\scriptstyle 1\\\scriptstyle 1
  \end{array}
  \end{array}
$, \.\, then\,\,\, $ sS_2\ ^t s=
  \begin {array}[t] {cc@{\extracolsep{0.2ex}}c}&\left (
  \begin {array}{ccccc}
  0& I^{(s-1)}&0&0\\
  I^{(s-1)}&0&0&0\\
  0&0&0&0\\
  0&0&0&0
  \end{array} \right)&\begin {array}{cc} \scriptstyle s-1\\\scriptstyle s-1\\\scriptstyle 1\\\scriptstyle 1\end{array}
  \end{array}
$. \end{center} Since $m\neq P_0$, therefore, there are $u_1, u_2\in
m\setminus P_0$ such that $m=s\oplus \langle u_1, u_2\rangle$ and
$$
\begin {array}[t] {cc@{\extracolsep{0.2ex}}c}\left (
  \begin {array}{ccccc}
Q\\
   v_0\\
  e_{2\nu +1}\\
  u_1\\
  u_2
  \end{array} \right)
  \end{array}S_2
{\begin {array}[t] {cc@{\extracolsep{0.2ex}}c}\
\begin{array}{l}^t\\ \\ \\ \\ \end{array}\left (
  \begin {array}{ccccc}
  Q\\
  v_0\\
  e_{2\nu +1}\\
  u_1\\
  u_2
  \end{array} \right)
  \end{array}}=
  \begin {array}[t] {cc@{\extracolsep{0.2ex}}c}&\left (
  \begin {array}{cccccccc}
  0& I^{(s-1)}&0&0& \ast &  \ast   \\
  I^{(s-1)}&0&0&0& \ast &  \ast\\
  0&0&0&1&0&0\\
  0&0&1&0&0&0\\
  \ast &  \ast&0&0&0&1\\
  \ast &  \ast&0&0&1&1
  \end{array} \right)
  \end{array}\,\,\,\,\,(\ast)
$$
Let $e_T=\langle v_0,
  e_{2\nu +1}, u_1, u_2\rangle$. Form $(\ast  )$ we deduce that $e_T$\, is a subspace of type\, (4,4,1,1)\, and
\,$e_T\cap P_0 = \langle\nu_0, e_{2\nu +1} \rangle$. Therefore $e_T$
is a encoding rule of transmitter and satisfying $s+e_T=m$.\

If $s'$ is another source state contained in $m$, then $s'\subset m,
P_0$, i.e., $s'\subset m\cap P_0=s$. While dim$s'$=dim$s$, so
$s'$=$s$, i.e., $s$ is the uniquely source state contained in $m$.\

Assuming the transmitter's encoding rules and the receiver's
decoding rules are chosen according to a uniform probability
distribution, we can assume that $\langle v_0,
  e_{2\nu +1}\rangle=\langle e_1,
  e_{2\nu +1}\rangle $, then $\langle v_0,
  e_{2\nu +1}\rangle^\perp =\langle e_1,e_2,\cdots, e_\nu,
  e_{\nu +2},\cdots
  e_{2\nu},
  e_{2\nu +1}\rangle $.

Let $n_1$ denote the number of subspaces of type $(2s,
2(s-1),s-1,1)$ contained in $\langle\nu_0, e_{2\nu +1} \rangle^\bot
$, and containing $ \langle\nu_0, e_{2\nu +1} \rangle$; $n_2$, the
number of subspaces of type $(m_0, 2s_0,s_0,1)$ contained in
$\langle\nu_0, e_{2\nu +1} \rangle^\bot $, and containing a fixed
subspace of type $(2s, 2(s-1),s-1,1)$ as above; and $n_3$, the
number of subspaces of type $(m_0, 2s_0,s_0,1)$ contained in
$\langle\nu_0, e_{2\nu +1} \rangle^\bot $, and containing $
\langle\nu_0, e_{2\nu +1} \rangle$.

{\bf Lemma 2.} (1) $n_1=N(2s-2, s-1; 2\nu -2)$; \

(2) $n_2=N(m_0-2s, s_0-s+1; 2(\nu -s))$; \

(3) $n_3=N(m_0-2, s_0; 2\nu -2)$. \

 Where $N(m, s; n )$ is the
number of subspaces of type $(m,s)$ in the $n$-dimensional
symplectic space $F_q^{(n)}$.

{\bf Proof.}  (1) We can assume that  $s$ is a subspace of type
\,$(2s, 2(s-1),s-1,1)$\, and\, $ \langle\nu_0, e_{2\nu +1}
\rangle\subset s\subset \langle\nu_0, e_{2\nu +1} \rangle^\bot $.
Clearly, $s$ has a form as follows\
$$
  \begin {array}[t] {cc@{\extracolsep{0.2ex}}c}s=&\left (
  \begin {array}{cccccccc}
  1&0&0&0&0&0&0&0\\
  0&0&0&0&0&0&1&0\\
  0&P_2&P_3&0&P_5&P_6&0&0
  \end{array} \right)&\begin {array}{cc} \scriptstyle 1\\\scriptstyle 1\\\scriptstyle 2s-2\\\end{array}
  \\&\begin {array}{cccccccc}\scriptstyle 1&\scriptstyle s-1& \scriptstyle \nu-s& \scriptstyle 1&\scriptstyle s-1&\scriptstyle \nu-s&
  \scriptstyle 1&\scriptstyle 1\end{array}&
  \end{array},
$$
where $(P_2,P_3,P_5,P_6)$ is a subspace of type $(2s-2,s-1)$ in the
symplectic space $F_q^{(2\nu -2)}$. Therefore,
$n_1=N(2s-2, s-1; 2\nu -2)$.\

(2) Assume that $P$ is a subspace of type $(m_0, 2s_0,s_0,1)$
containing a fixed subspace of type $(2s, 2(s-1),s-1,1)$ as above
and $P\subset\langle\nu_0, e_{2\nu +1} \rangle^\bot$. It is easy to
know that $P$ has a form as follows
$$
  \begin {array}[t] {cc@{\extracolsep{0.2ex}}c}P=&\left (
  \begin {array}{cccccccc}
  1&0&0&0&0&0&0&0\\
  0&0&0&0&0&0&1&0\\
  0& I^{(s-1)}&0&0&0&0&0&0\\
  0&0&0&0& I^{(s-1)}&0&0&0\\
  0&0&L_3&0&0&L_6&0&0
  \end{array} \right)&\begin {array}{cc} \scriptstyle 1\\\scriptstyle 1\\\scriptstyle s-1\\\scriptstyle s-1\\\scriptstyle m_0-2s\end{array}
  \\&\begin {array}{cccccrcc}\scriptstyle 1\,\,\,&\scriptstyle s-1\,\,\,& \scriptstyle \nu-s\,\,& \scriptstyle 1\,\,\,&\scriptstyle s-1\,\,\,&\scriptstyle \nu-s&\,\
  \scriptstyle 1&\scriptstyle 1\end{array}&
  \end{array},
$$
where $(L_3,L_6)$ is a subspace of type $(m_0-2s, s_0-s+1)$ in the
symplectic space $F_q^{2(\nu -s)}$. Therefore,
$n_2=N(m_0-2s, s_0-s+1; 2(\nu -s))$.\

(3) Similar to the proof of (1), we have  $n_3=N(m_0-2, s_0; 2\nu
-2)$.\

{\bf Lemma 3.} The number of the source states is $|S|=N(2s-2, s-1;
2\nu -2)N(m_0-2s, s_0-s+1; 2(\nu -s))/N(m_0-2, s_0; 2\nu -2)$.\

{\bf Proof.}\,  $|S|$ is the number of subspace of type $(2s,
2(s-1),s-1,1)$ contained in $P_0$, and containing $ \langle\nu_0,
e_{2\nu +1} \rangle$. In order to compute $|S|$, we define a
(0,1)-matrix, whose rows are indexed by the subspaces of type $(2s,
2(s-1),s-1,1)$ containing $\langle\nu_0, e_{2\nu +1} \rangle$ and
contained in $\langle\nu_0, e_{2\nu +1} \rangle^\bot$ whose columns
are indexed by the subspaces of type $(m_0, 2s_0,s_0,1)$ containing
$\langle\nu_0, e_{2\nu +1} \rangle$ and contained in $\langle\nu_0,
e_{2\nu +1} \rangle^\bot$, and with a 1 or 0 in the (i,j) position
of the matrix, if the i-th subspace of type $(2s, 2(s-1),s-1,1)$ is
or is not contained in the j-th subspace of type $(m_0,
2s_0,s_0,1)$, respectively. If we count the number of 1's in the
matrix by rows, we get $n_1\cdot n_2$, where $n_1$ is the number of
rows and $n_2$ is the number of 1's in each row. If we count the
number of 1's in the matrix by columns, we get $n_3\cdot |S|$, where
$n_3$ is the number of columns and $|S|$ is the number of 1's in
each column. Thus we have $n_1\cdot n_2=n_3\cdot |S|$.

 {\bf Lemma 4.} The number of the encoding rules of transmitter is
 $|E_T|=q^{4(\nu -1)}$.\

{\bf Proof.}\, Since $e_T$ is a subspace of type (4,4,1,1) and
$e_T\cap P_0 = \langle\nu_0, e_{2\nu +1} \rangle$, the transmitter's
encoding rules have the form as follows
$$
 \begin {array}[t] {cc@{\extracolsep{0.2ex}}c}e_T=&\left (
  \begin {array}{cccccccc}
  1&0&0&0&0&0&0&0\\
  0&0&0&0&0&0&1&0\\
  0&R_2&R_3&1&R_5&R_6&0&0\\
  0&L_2&L_3&0&L_5&L_6&0&1
  \end{array} \right)&\begin {array}{cc} \scriptstyle 1\\\scriptstyle 1\\\scriptstyle 1\\\scriptstyle 1\\\end{array}
  \\&\begin {array}{cccccccc}\scriptstyle 1&\scriptstyle s-1& \scriptstyle \nu-s& \scriptstyle 1&\scriptstyle s-1&\scriptstyle\nu-s&
  \scriptstyle 1&\scriptstyle 1\end{array}&
  \end{array},
$$
where $R_2,R_3,R_5,R_6, L_2,L_3,L_5,L_6$ arbitrarily. Therefore,
$|E_T|=q^{4(\nu -1)}$.\

{\bf Lemma 5.} The number of the decoding rules of receiver is
 $|E_R|=q^{2\nu }$.\

{\bf Proof.} Since $ e_R$\, is a subspace of type\, (2,2,0,1) in the
$(2\nu +2)$-dimensional pseudo-symplectic space
$F_q^{(2\nu +2)}$,  it has the form as follows
$$
  \begin {array}[t] {cc@{\extracolsep{0.2ex}}c}e_R=&\left (
  \begin {array}{cccccccc}
  0&0&0&0&0&0&1&0\\
  R_1&R_2&R_3&R_4&R_5&R_6&0&1
  \end{array} \right)&\begin {array}{cc} \scriptstyle 1\\\scriptstyle 1\\\end{array}
  \\\begin {array}{cc} \end{array}
 \ &\begin {array}{cccccccc}\scriptstyle 1\,\,&\scriptstyle s-1& \scriptstyle \nu-s\,& \scriptstyle 1\,&\scriptstyle s-1&\scriptstyle \nu-s\,\,&
  \scriptstyle 1\,\,&\scriptstyle 1\end{array}&
  \end{array},
$$where $R_1, R_2,R_3,R_4,R_5,R_6$ arbitrarily. Therefore,
$|E_R|=q^{2\nu }$.\

{\bf Lemma 6.} For any $m\in M$, let the number of $ e_T$ and $ e_R$
contained in $m$ be $a$ and $b$, respectively. Then $a=q^{4(s-1)}$,
$b=q^{2s}$.\

{\bf Proof.} Let $m$ be a message.  From the definition of $m$, we
may take $m$ as follows:
$$
 \begin {array}[t] {cc@{\extracolsep{0.2ex}}c}m=&\left (
  \begin {array}{cccccccc}
  1&0&0&0&0&0&0&0\\
  0&0&0&0&0&0&1&0\\
  0& I^{(s-1)}&0&0&0&0&0&0\\
  0&0&0&0& I^{(s-1)}&0&0&0\\
  0&0&0&1&0&0&0&0\\
  0&0&0&0&0&0&0&1
  \end{array} \right)&\begin {array}{cc} \end{array}
  \\&\begin {array}{cccccrcc}\scriptstyle 1&\scriptstyle s-1\,\,& \scriptstyle \nu-s\,& \scriptstyle 1\,\,&\scriptstyle s-1\,\,&\scriptstyle \nu-s&
  \scriptstyle 1&\scriptstyle 1\end{array}&
  \end{array},
$$
If $e_T\subset  m$, then we can assume
$$
 \begin {array}[t] {cc@{\extracolsep{0.2ex}}c}e_T=&\left (
  \begin {array}{cccccccc}
  1\,&0\,&0\,\,&0\,&0\,&0\,\,&0\,&0\\
  0&0&0&0&0&0&1&0\\
  0&R_2&0&1&R_5&0&0&0\\
  0&L_2&0&0&L_5&0&0&1
  \end{array} \right)&\begin {array}{cc} \scriptstyle 1\\\scriptstyle 1\\\scriptstyle 1\\\scriptstyle 1\\\end{array}
  \\&\begin {array}{cccccccc}\scriptstyle 1&\scriptstyle s-1&\scriptstyle \nu-s& \scriptstyle 1&\scriptstyle s-1&\scriptstyle\nu-s&
  \scriptstyle 1&\scriptstyle 1\end{array}&
  \end{array},
$$
where $R_2,R_5, L_2, L_5 $ arbitrarily. Therefore, $a=q^{4(s
-1)}$.\\
If $e_R\subset  m$, then we can assume
$$
  \begin {array}[t] {cc@{\extracolsep{0.2ex}}c}e_R=&\left (
  \begin {array}{cccccccc}
  0&0&0&0&0&0&1&0\\
  R_1&R_2&0&R_4&R_5&0&0&1
  \end{array} \right)&\begin {array}{cc} \scriptstyle 1\\\scriptstyle 1\\\end{array}
  \\\begin {array}{cc} \end{array}
 \ &\begin {array}{cccccccc}\scriptstyle 1&\scriptstyle s-1& \scriptstyle \nu-s& \scriptstyle 1&\scriptstyle s-1&\scriptstyle \nu-s&
  \scriptstyle 1&\scriptstyle 1\end{array}&
  \end{array},
$$where $R_1,R_2,R_4, R_5$ arbitrarily. Therefore,
$b=q^{2s }$.\

{\bf Lemma 7.} The number of the messages is $|M|=q^{4(\nu
-s)}|S|$.\

{\bf Proof.} We know that a message contains a source state and the
number of the transmitter's encoding rules contained in a message is
$a$. Therefore we have  $|M|=|S||E_T|/a=q^{4(\nu -s)}|S|$.\

{\bf Lemma 8.} (1) For any $e_T\in E_T$, the number of $e_R$ which
is incidence with $e_T$ is $c=q^2$.\

(2) For any $e_R\in E_R$, the number of $e_T$ which is incidence
with $e_R$ is $d=q^{2(\nu -1)}$.\

{\bf Proof.} (1) Assume that $e_T\in E_T$, $e_T$ is a subspace of
type (4,4,1,1) and $e_T\cap P_0 = \langle\nu_0, e_{2\nu +1}
\rangle$, we may take $e_T$ as follows:
$$
 \begin {array}[t] {cc@{\extracolsep{0.2ex}}c}e_T=&\left (
  \begin {array}{cccccccc}
  1\,\,&0\,\,&0\,\,&0\,\,&0\,\,&0\,\,\,&0\,\,&0\\
  0\,\,&0\,\,&0\,\,&0\,\,&0\,\,&0\,\,\,&1\,\,&0\\
  0\,\,&0\,\,&0\,\,&1\,\,&0\,\,&0\,\,\,&0\,\,&0\\
  0\,\,&0\,\,&0\,\,&0\,\,&0\,\,&0\,\,\,&0\,\,&1
  \end{array} \right)&\begin {array}{cc} \scriptstyle 1\\\scriptstyle 1\\\scriptstyle 1\\\scriptstyle 1\\\end{array}
  \\&\begin {array}{cccccccc}\scriptstyle 1&\scriptstyle s-1&\scriptstyle \nu-s& \scriptstyle 1&\scriptstyle s-1&\scriptstyle\nu-s&
  \scriptstyle 1&\scriptstyle 1\end{array}&
  \end{array}.
$$
If $e_R\subset e_T$, then we can assume
$$
  \begin {array}[t] {cc@{\extracolsep{0.2ex}}c}e_R=&\left (
  \begin {array}{cccccccc}
  0\,&0\,&0\,&0\,&0\,&0\,&1\,&0\\
  R_1\,&0\,&0&R_4\,&0\,&0\,&0\,&1
  \end{array} \right)&\begin {array}{cc} \scriptstyle 1\\\scriptstyle 1\\\end{array}
  \\\begin {array}{cc} \end{array}
 \ &\begin {array}{cccccccc}\scriptstyle 1&\scriptstyle s-1& \scriptstyle \nu-s& \scriptstyle 1&\scriptstyle s-1&\scriptstyle \nu-s&
  \scriptstyle 1&\scriptstyle 1\end{array}&
  \end{array},
$$where $R_1,R_4,$ arbitrarily. Therefore,
$c=q^2$.\

(2) Assume that  $ e_R\subset E_R$, $e_R$ is a subspace of type\,
(2,2,0,1) in the $(2\nu +2)$-dimensional pseudo-symplectic space
$F_q^{(2\nu +2)}$,  we may take $e_R$ as follows:
$$
 \begin {array}[t] {cc@{\extracolsep{0.2ex}}c}e_R=&\left (
  \begin {array}{cccccccc}
  0\,\,&0\,\,&0\,\,&0\,\,&0\,\,&0\,\,\,&1\,\,&0\\
  0\,\,&0\,\,&0\,\,&0\,\,&0\,\,&0\,\,\,&0\,\,&1
  \end{array} \right)&\begin {array}{cc} \scriptstyle 1\\\scriptstyle 1\\\end{array}
  \\&\begin {array}{cccccccc}\scriptstyle 1&\scriptstyle s-1&\scriptstyle \nu-s& \scriptstyle 1&\scriptstyle s-1&\scriptstyle\nu-s&
  \scriptstyle 1\,\,&\scriptstyle 1\end{array}&
  \end{array}.
$$
If $e_T\supset e_R$, then we can assume
$$
 \begin {array}[t] {cc@{\extracolsep{0.2ex}}c}e_T=&\left (
  \begin {array}{cccccccc}
  1&0&0&0&0&0&0&0\\
  0&0&0&0&0&0&1&0\\
  0&R_2&R_3&1&R_5&R_6&0&0\\
  0&0&0&0&0&0&0&1
  \end{array} \right)&\begin {array}{cc} \scriptstyle 1\\\scriptstyle 1\\\scriptstyle 1\\\scriptstyle 1\\\end{array}
  \\&\begin {array}{cccccccc}\scriptstyle 1&\scriptstyle s-1&\scriptstyle \nu-s& \scriptstyle 1&\scriptstyle s-1&\scriptstyle\nu-s&
  \scriptstyle 1&\scriptstyle 1\end{array}&
  \end{array},
$$
where $R_2,R_3, R_5, R_6 $ arbitrarily. Therefore, $d=q^{2(\nu
-1)}$.

{\bf Lemma 9.} For any  $m\in M$ and $e_R\subset m$, the number of
$e_T$ contained in $m$ and containing $e_R$ is $q^{2(s -1)}$ .\

{\bf Proof.} The matrix of $m$ is like lemma 6, then for any
$e_R\subset m$, assume that
$$
  \begin {array}[t] {cc@{\extracolsep{0.2ex}}c}e_R=&\left (
  \begin {array}{cccccccc}
  0&0&0&0&0&0&1&0\\
  R_1&R_2&0&R_4&R_5&0&0&1
  \end{array} \right)&\begin {array}{cc} \scriptstyle 1\\\scriptstyle 1\\\end{array}
  \\\begin {array}{cc} \end{array}
 \ &\begin {array}{cccccccc}\scriptstyle 1&\scriptstyle s-1& \scriptstyle \nu-s& \scriptstyle 1&\scriptstyle s-1&\scriptstyle \nu-s&
  \scriptstyle 1&\scriptstyle 1\end{array}&
  \end{array},
$$ if $e_T\subset m$ and $e_T\supset e_R$, then $e_T$ has a form as
follows$$
 \begin {array}[t] {cc@{\extracolsep{0.2ex}}c}e_T=&\left (
  \begin {array}{cccccccc}
  1\,&0\,&0\,\,&0\,&0\,&0\,\,&0\,&0\\
  0&0&0&0&0&0&1&0\\
  0&L_2&0&1&L_5&0&0&0\\
  0&R_2&0&0&R_5&0&0&1
  \end{array} \right)&\begin {array}{cc} \scriptstyle 1\\\scriptstyle 1\\\scriptstyle 1\\\scriptstyle 1\\\end{array}
  \\&\begin {array}{cccccccc}\scriptstyle 1&\scriptstyle s-1&\scriptstyle \nu-s& \scriptstyle 1&\scriptstyle s-1&\scriptstyle\nu-s&
  \scriptstyle 1&\scriptstyle 1\end{array}&
  \end{array},
$$
where $L_2, L_5$ arbitrarily. Therefore, the number of $e_T$
contained in $m$ and containing $e_R$ is $q^{2(s -1)}$.\

{\bf Lemma 10.} Assume that $m_1$ and $m_2$ are two distinct
messages which commonly contain a transmitter's encoding rule
$e_T'$. $s_1$ and $s_2$ contained in $m_1$ and $m_2$ are two source
states, respectively. Assume that $s_0=s_1\cap s_2$, dim $s_0=k$,
then $2\leq k\leq 2s-1$, and\

(1) The number of $e_R$ contained in $m_1\cap m_2$ is $q^k$; \

(2) For any $e_R\subset m_1\cap m_2$, the number of $e_T$ contained
in $m_1\cap m_2$ and containing $e_R$ is $q^{k-2}$.

{\bf Proof.} Since $m_1=s_1+e_T', m_2=s_2+e_T'$ and $m_1\neq m_2$,
then $s_1\neq  s_2$. And because of $ \langle\nu_0, e_{2\nu +1}
\rangle\subset s_1, s_2$, therefore, $2\leq k\leq 2s-1$.\

(1) Assume that $s'_i$ is the complementary subspace of $s_0$ in the
$s_i$, then $s_i=s_0+s'_i\, \,\,(i=1,2)$. From
$m_i=s_i+e_T'=s_0+s'_i+e_T'$ and $\ s_{i}=m_{i}\cap P_0$ $(i=1,2)\
,$ we have $s_{0}=\left(m_{1}\cap P_0\right)\bigcap \left(m_{2}\cap
P_0\right) =m_{1}\cap m_{2}\cap P_0 = s_{1}\cap m_{2}=s_{2}\cap
m_{1}$ and $m_{1}\cap m_{2}=( s_{1}+ e^{'}_{T})\cap m_{2} =(
s_{0}+s^{'}_{1}+ e^{'}_{T})\cap m_{2} =((s_{0}+
e^{'}_{T})+s^{'}_{1})\cap m_{2}\ .$\ Because $s_{0}+
e^{'}_{T}\subset m_{2}\ ,m_{1}\cap m_{2}= (s_{0}+ e^{'}_{T} )+ (
s^{'}_{1}\cap m_{2} )\ .$  While $s^{'}_{1}\cap m_{2}\subseteq
s_{1}\cap m_{2}=s_{0}\ ,$ $m_{1}\cap m_{2}=s_{0}+ e^{'}_{T}\ .$
Therefore dim $(m_1\cap m_2)=k+2$. From the definition of the
message, we may take $m_1$ and $m_2$ as follows respectively
$$
\begin {array}[t] {cc@{\extracolsep{0.2ex}}c}m_1=&\left (
  \begin {array}{cccccccc}
  1\,&0\,&0\,&0\,&0\,&0\,\,&0\,&0\\
  0\,&0\,&0\,&0\,&0\,&0\,\,&1\,&0\\
  0\,& A_2\,&0\,&0\,&A_5\,&0\,\,&0\,&0\\
  0\,&A'_2\,&0\,&0\,& A'_5\,&0\,\,&0\,&0\\
  0\,&0\,&0\,&1\,&0\,&0\,\,&0\,&0\\
  0\,&0\,&0\,&0\,&0\,&0\,\,&0\,&1
  \end{array} \right)&\begin {array}{cc}\scriptstyle 1\\\scriptstyle 1\\\scriptstyle s-1\\\scriptstyle s-1\\\scriptstyle 1\\\scriptstyle 1 \end{array}
  \\&\begin {array}{cccccccc}\scriptstyle 1&\scriptstyle s-1&\scriptstyle \nu-s& \scriptstyle 1&\scriptstyle s-1&\scriptstyle\nu-s&
  \scriptstyle 1&\scriptstyle 1\end{array}&
  \end{array},$$

 $$ \begin {array}[t] {cc@{\extracolsep{0.2ex}}c}m_2=&\left (
  \begin {array}{cccccccc}
   1\,&0\,&0\,&0\,&0\,&0\,\,&0\,&0\\
  0\,&0\,&0\,&0\,&0\,&0\,\,&1\,&0\\
  0\,& B_2\,&0\,&0\,&B_5\,&0\,\,&0\,&0\\
  0\,&B'_2\,&0\,&0\,& B'_5\,&0\,\,&0\,&0\\
  0\,&0\,&0\,&1\,&0\,&0\,\,&0\,&0\\
  0\,&0\,&0\,&0\,&0\,&0\,\,&0\,&1
  \end{array} \right)&\begin {array}{cc}\scriptstyle 1\\\scriptstyle 1\\\scriptstyle s-1\\\scriptstyle s-1\\\scriptstyle 1\\\scriptstyle 1 \end{array}
  \\&\begin {array}{cccccccc}\scriptstyle 1&\scriptstyle s-1&\scriptstyle \nu-s& \scriptstyle 1&\scriptstyle s-1&\scriptstyle\nu-s&
  \scriptstyle 1&\scriptstyle 1\end{array}&
  \end{array}.
$$
Thus
$$
\begin {array}[t] {cc@{\extracolsep{0.2ex}}c}m_1\cap m_2=&\left (
  \begin {array}{cccccccc}
   1\,&0\,&0\,&0\,&0\,&0\,\,&0\,&0\\
  0\,&0\,&0\,&0\,&0\,&0\,\,&1\,&0\\
  0\,& P_2\,&0\,&0\,&P_5\,&0\,\,&0\,&0\\
  0\,&P'_2\,&0\,&0\,& P'_5\,&0\,\,&0\,&0\\
  0\,&0\,&0\,&1\,&0\,&0\,\,&0\,&0\\
  0\,&0\,&0\,&0\,&0\,&0\,\,&0\,&1
  \end{array} \right)&\begin {array}{cc}\scriptstyle 1\\\scriptstyle 1\\\scriptstyle s-1\\\scriptstyle s-1\\\scriptstyle 1\\\scriptstyle 1\end{array}
  \\&\begin {array}{cccccccc}\scriptstyle 1&\scriptstyle s-1&\scriptstyle \nu-s& \scriptstyle 1&\scriptstyle s-1&\scriptstyle\nu-s&
  \scriptstyle 1&\scriptstyle 1\end{array}&
  \end{array}.$$
and
\begin{center}
dim$\begin {array}[t] {cc@{\extracolsep{0.2ex}}c}&\left (
  \begin {array}{cccccccc}
  0& P_2&0&0&P_5&0&0&0\\
  0&P'_2&0&0& P'_5&0&0&0
  \end{array} \right)&\begin {array}{cc}\end{array}
  \\&\begin {array}{cccccccc}\end{array}&
  \end{array}=k-2.$
\end{center}
If for any $e_R\subset m_1\cap m_2$, then
$$
  \begin {array}[t] {cc@{\extracolsep{0.2ex}}c}e_R=&\left (
  \begin {array}{cccccccc}
  0&0&0&0&0&0&1&0\\
  R_1&R_2&0&R_4&R_5&0&0&1
  \end{array} \right)&\begin {array}{cc} \scriptstyle 1\\\scriptstyle 1\\\end{array}
  \\\begin {array}{cc} \end{array}
 \ &\begin {array}{cccccccc}\scriptstyle 1&\scriptstyle s-1& \scriptstyle \nu-s& \scriptstyle 1&\scriptstyle s-1&\scriptstyle \nu-s&
  \scriptstyle 1&\scriptstyle 1\end{array}&
  \end{array},
$$
where $R_1, R_4$ arbitrarily, and every row of $(0 \,\,R_2\,\,
0\,\, 0\,\, R_5\,\, 0\,\, 0 \,\,0)$ is the linear combination of
the base of $$\left (
  \begin {array}{cccccccc}
  0& P_2&0&0&P_5&0&0&0\\
  0&P'_2&0&0& P'_5&0&0&0
  \end{array} \right).$$ So it is easy to know that the number of $e_R$ contained in $m_1\cap m_2$ is
  $q^k$.\

  (2) Assume that $ m_1\cap m_2$ has the form of (1), then for any $e_R\subset m_1\cap
  m_2$, we can assume that
$$
  \begin {array}[t] {cc@{\extracolsep{0.2ex}}c}e_R=&\left (
  \begin {array}{cccccccc}
  0&0&0&0&0&0&1&0\\
  R_1&R_2&0&R_4&R_5&0&0&1
  \end{array} \right)&\begin {array}{cc} \scriptstyle 1\\\scriptstyle 1\\\end{array}
  \\\begin {array}{cc} \end{array}
 \ &\begin {array}{cccccccc}\scriptstyle 1&\scriptstyle s-1& \scriptstyle \nu-s& \scriptstyle 1&\scriptstyle s-1&\scriptstyle \nu-s&
  \scriptstyle 1&\scriptstyle 1\end{array}&
  \end{array},$$
If $e_T\subset m_1\cap m_2$ and $e_R\subset e_T $, then $e_T $ has
the form as follows
$$
 \begin {array}[t] {cc@{\extracolsep{0.2ex}}c}e_T=&\left (
  \begin {array}{cccccccc}
  1\,&0\,&0\,\,&0\,&0\,&0\,\,&0\,&0\\
  0&0&0&0&0&0&1&0\\
  0&L_2&0&1&L_5&0&0&0\\
  0&R_2&0&0&R_5&0&0&1
  \end{array} \right)&\begin {array}{cc} \scriptstyle 1\\\scriptstyle 1\\\scriptstyle 1\\\scriptstyle 1\\\end{array}
  \\&\begin {array}{cccccccc}\scriptstyle 1&\scriptstyle s-1&\scriptstyle \nu-s& \scriptstyle 1&\scriptstyle s-1&\scriptstyle\nu-s&
  \scriptstyle 1&\scriptstyle 1\end{array}&
  \end{array},
$$
where every row of $(0 \,\,L_2\,\, 0\,\, 0\,\, L_5\,\, 0\,\, 0
\,\,0)$ is the linear combination of the base of $$\left (
  \begin {array}{cccccccc}
  0& P_2&0&0&P_5&0&0&0\\
  0&P'_2&0&0& P'_5&0&0&0
  \end{array} \right),$$
 then the number of $e_T$ contained
in $m_1\cap m_2$ and  containing $e_R$ is $q^{k-2}$.

{\bf Theorem 1.} The parameters of constructed authentication codes
with arbitration are
$$|S|=N(2s-2, s-1; 2\nu -2)N(m_0-2s, s_0-s+1; 2(\nu -s))/N(m_0-2,
s_0; 2\nu -2);$$
$$|M|=q^{4(\nu -s)}|S|;\,\,\,\,\,\,\,\,\,\,\,\,\,\, |E_T|=q^{4(\nu -1)};\,\,\,\,\,\,\,\,\,\,\,\,\,\, |E_R|=q^{2\nu }.
$$

{\bf Theorem 2.} In the ${A}^{2} $ authentication codes, if the
transmitter's encoding rules and the receiver's decoding rules are
chosen according to a uniform probability distribution, the largest
probabilities of success for different types of deceptions:
$$
P_I= \frac{1}{q^{2(\nu -s)}};\,\,\,\,\,\,P_S=
\frac{1}{q};\,\,\,\,\,\,P_T= \frac{1}{q};\,\,\,\,\,\,P_{R_0}=
\frac{1}{q^{2(\nu -s)}};\,\,\,\,\,\,P_{R_1}= \frac{1}{q};
$$

{\bf Proof. } \ (1)\ The number of the transmitter's encoding rules
contained in a message is $b,$ then
$$P_{I}=\max\limits_{m\in M}\left\{\frac{\mid{e_{R}\in E_{R}|e_{R}
\subset m}\mid}{\mid E_{R} \mid}\right\}= \frac{b}{\mid
E_{R}\mid}=\frac{1}{q^{2(\nu -s)}} .$$

 (2) \ Assume that opponent
get $\ m_{1}$ which is from transmitter, and send $m_{2}$\ instead
of $\ m_{1}$, when $\ s_{1}$ contained in $\ m_{1}$ is different
from $\ s_{1}$ contained in $\ m_{2}$, the opponent's substitution
attack can success. Because $\ e_{R}\subset \ e_{T}\subset m_{1}$,
thus the opponent select $\ e_{T}^{'}\subset m_{1}$ , satisfying $\
m_{2}=s_{2}+e_{T}^{'}$\ and $\ {\rm dim}(s_{1}\bigcap s_{2} )=k$,
then
$$P_{S}=\max\limits_{m\in M}\left\{\frac{\max\limits_{m \neq m^{'}\in M} \mid {e_{R}\in
E_{R}|e_{R}\subset m \ \rm and \ e_{R}\subset m^{'}} \mid} {\mid
{e_{R}\in E_{R}|e_{R}\subset m}\mid} \right\}=\frac{q^k}{b}$$ where
$k=2s-1, P_s= \frac{1}{q}$ is the largest.\

(3) Let $e_T$ be the transmitter's secret encoding rules, $s$ be a
source state, and $m_1$ be the message corresponding to the source
state $s$ encoded by $e_T$. Then the number of the receiver's
decoding rules contained in $m_1$ is $e_R$. Assume that $m_2$ is a
distinct message corresponding to $s$, but $m_2$ cannot be encoded
by $e_T$. Then $m_1\cap m_2$ contains $q$ receiver's decoding rules
at most. Therefore the probability of transmitter's successful
impersonation attack is
 $$P_{T}=\max\limits_{e_{T}\in E_{T}}\left\{\frac{\max\limits_{m\in M ,
 e_{T}\not\subset m}\mid\{e_{R}\in E_{R}|e_{R}\subset m
\cap e_{T}\}\mid}{\mid\{e_{R} \in E_{R}|e_{R}\subset
e_{T}\}\mid}\right\} =\frac{q}{q^2}=\frac{1}{q}$$.

(4) Let $e_R$ be the receiver's decoding rule, we have known that
the number of transmitter's encoding rules containing $e_R$ is
$q^{2(\nu-s)}$ and a message containing $e_R$ has $q^{2(s-1)}$
transmitter's encoding rules. Hence the probability of a receiver's
successful impersonation attack is
 $$P_{R_0}=\max\limits_{e_{R}\in E_{R}}\left\{\frac{\max\limits_{m\in M
}\mid\{e_{T}\in E_{T}|e_{T}\subset m {\rm \ and} \ e_{R}\subset
e_{T}\mid}{\mid\{e_{T} \in E_{T}| e_{R}\subset
e_{T}\mid}\right\}=\frac{q^{2(s-1)}}{q^{2(\nu
-1)}}=\frac{1}{q^{2(\nu -s)}}.$$

(5)\ Assume that the receiver declares to receive a message $m_{2}$
instead of $m_{1},$ when $s_{2}$ contained in $m_{1}$ is different
from $s_{2}$ contained in $m_{2}$, the receiver's substitution
attack can be successful. Since $e_{R}\subset e_{T}\subset m_{1}$,
receiver is superior to select $e_{T}^{'}\ $,\ satisfying
$e_{R}\subset e_{T}^{'}\subset m_{1}$ , thus
$m_{2}=s_{2}+e_{T}^{'}$, and ${\rm dim}(s_{1}\cap s_{2})=k$ as large
as possible. Therefore, the probability of a receiver's successful
substitution attack is
$$P_{R_{1}}=\max\limits_{e_{R}\in E_{R},m\in M}\left\{\frac{\max\limits_{m^{'}
\in M
 }\mid\{e_{T}\in E_{T}|e_{T}\subset m ,m^{'}\ {\rm  and} \
e_{R}\subset e_{T}\}\mid}{\mid\{e_{T} \in E_{T}|e_{R}\subset
e_{T}\}\mid}\right\}=\frac{q^{k-2}}{q^{2(s-1)}},$$ where $k=2s-1,\,
P_{R_{1}}= \frac{1}{q} $ is the largest.\\

\end{document}